# Planar Visibility Graph Network Algorithm For Two Dimensional Timeseries

Jie Liu[1,2,*] and Qingqing Li[1]
1) Research Centre of Nonlinear Science, Wuhan Textile University, Wuhan, P R, China, 430073
2) Mathematics Department, Wuhan Textile University, Wuhan, P R, China, 430073
E-mail: liujie_hch@163.com

**Abstract:** In this brief paper, a simple and fast computational method, the Planar Visibility Graph Networks Algorithm was proposed based on the famous Visibility Graph Algorithm, which can fulfill converting two dimensional timeseries into a planar graph. The constructed planar graph inherits several properties of the series in its structure. Thereby, periodic series, random series, and chaotic series convert into quite different networks with different average degree, characteristic path length, diameter, clustering coefficient, different degree distribution, and modularity, etc. By means of this new approach, with such different networks measures, one can characterize two dimensional timeseries from a new viewpoint of complex networks.

## 1. Some backgrounds.

Disregarding any underlying process, no matter it is a physical, chemical, or an economical processes, one can consider a time series as an ordered data set of real values and deal with it by some naive mathematical games of translating this set into a related but looks quite different mathematical object with the aids of an abstract map. Researchers always concern that, after mapping timeseries into other forms, what will happen? Which properties of the original timeseries will be conserved? Is there some other transformation strategies? what usage of the proposed mathematical representations for this timeseries will be found for further applications [1]. As mentioned by Angel M. Nunez et al in [2], [3], the preceding mathematical game had attracted unexpected practical interests as it opens a possibility of analyzing a time series from an alternative viewpoint. Of course, the information hidden in the original dataset (timeseries) should be somehow conserved in the constructed map. This motivation is completed when the new formulation belongs to a relatively mature mathematical research field, where information encoded in such a translation can be effectively disentangled and processed. This can be seen as basical motivation for mapping time series into networks or duality analysis of timeseries and networks. There are some new approaches published in this direction (see [4],[5], and references therein), for example, different undirected or directed visibility algorithms proposed by Lucas Lacasa et al.[2],[3], the recurrence networks constructed from a timeseries proposed by J Kurth et al.[6],[7],[8], mapped each cycle of a pseudo-periodic time series into a node in a graph proposed by Zhang & Small [9], [10], embedding time series in an potential phase space and set each point corresponded to a node to construct a network proposed by Xu & Small [11], etc. Very recently, Donner et al., had presented a technique in which the recurrence matrix obtained by imposing a threshold in the minimum distance between two points in the phase space was used based on the properties of recurrence in the reconstructed phase space of a underlying dynamical system [6]-[8]. They mapped timeseries into networks (an undirected, unweighted graph) based on the recurrence adjacency matrix. Furthermore, Campanharo et al., discussed duality of timeseries and networks [12]. This idea can be seen as created along with the original ideas of Shirazi et al.[13], Strozzi et al.[14] and Haraguchi et al.[15] of a surjective mapping which admits an inverse operation. This approach began the reciprocal possibility of benefiting from time series analysis to study the structure and properties of networks. These methods had been used for dectecting unstable periodic orbits in a chaotic attractor, differing chaotic, hyperchaotic, random and noisy periodic underlying dynamics from each other, and even it had been subsequently applied to characterize cardiac dynamics, economics market, two-phase flows identification, etc, successfully [16]-[27].

Among all these methods of mapping, the Visibility Graph (VG) methods proposed by Lacasa et al.[2] might had attracted most attentions in theory analysis and application designing fields because of its intrinsic non-locality, its low computational cost, its straightforward implementation and its quite 'simple' rules of inherit the timeseries properties in the structure of the associated networks. These features are going to make it easier to find connections between the underlying physical processes and the networks obtained from them by a direct analysis of the latter. A short review on this direction was recently published by Angel M. Nunez, Lucas Lacasa, Jose Patricio Gomez and Bartolo Luque recently [3]. In [3], the author had presented different versions of the VG algorithm along with its most notable properties, that in many cases can be derived analytically, and also several applications are addressed in their paper (one can consult them for more details).

Since new approaches, new paradigms to deal with nonlinear timeseries' complexity are not only welcome, but really needed. In this paper, we will further extended the VG algorithm to two dimension cases, since the VG algorithm and recurrence networks are mostly constructed for one dimension timeseries based on nonlinear correlation functions, embedding algorithms, multifractal spectra, projection theorems, etc. Some tools used by researches might increase the complexity parallel to the complexity of the process/series. Is there some more 'Simple' methods for high dimension timeseries directly? We will give some discussion on this topic. Our method can be seen as a new member of the family of visibility algorithms. It will constitute one of other possibilities to map a two dimension time series into a network and subsequently analyze the structure of the series through the set of tools developed in the complex network theory. It can also be seen as a natural bridge between complex networks theory and high dimensional timeseries analysis.

The rest of this paper is organized as follows: in section 2, the classical VG algorithms is recalled. In section 3, a new two dimension VG algorithm is proposed based on classical VG algorithm. In section 4, this new algorithm is used for differing two dimensional chaotic, random and periodic underlying dynamics from each other directly under the viewpoint of associated complex networks. A brief conclusion are concluded in the last section.

## 2. Classical visibility algorithms: a recall

The idea of mapping time series into networks (graphs) seems attractive because it bridges the gap between two prolific fields of modern science as nonlinear timeseries analysis and complex networks analysis. It has caught the attention of many research groups allover the world which have contributed to the topic with different strategies of mapping. While recall of such strategies is beyond the scope of this work, we shall only concern on one of the most attractive case, visibility algorithm proposed in [2]. In this section, we will firstly introduce the classical VG methods to map one dimension timeseries into a related networks.

**2.1 Classical visibility algorithm: definition**

The classical visibility graph (VG) algorithm [2] assigns each datum of the series to a node in the natural visibility graph. Assume that the time series for: $\{x_i\}$, i=1,⋯,N, the resulting complex network denoted as: G=<V,E>, here V=$\{v_i\}$, i=1,⋯,N is the set of nodes, E=$\{e_i\}$, i=1,⋯,N, is the connection matrix of the network. For ease of understanding, the classical VG algorithm is listed as follows (see figure 1 for an illustration):   If any of the i<n<j, both have

$$\frac{x_i - x_n}{i - n} > \frac{x_i - x_j}{i - j}$$

then the i node and the j node is visual, otherwise, invisible.

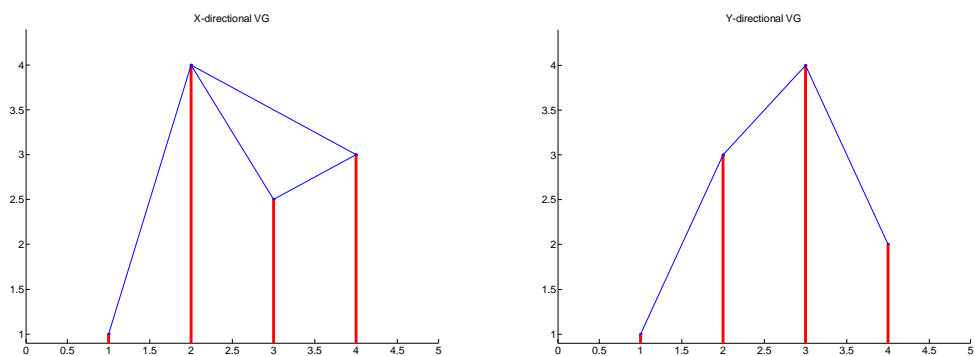

Figure 1. Sample for 1-D VG algorithm for two short timeseries: (a) Ts(1)= [1.0, 4.0, 2.5, 3.0] (Left), and (b) Ts(2) =[1.0, 3.0, 4.0, 2.0] (Right).

**2.2 Classical visibility algorithm: characters**

It can easily checked that by means of the present algorithm, the associated graph extracted from a timeseries behaves as follows [2],[3]:

(i) Connected: each node sees at least its nearest neighbors (left-hand side and right-hand side).

(ii) Undirected: the way the algorithm is built up, there is no direction defined in the links.

(iii) Invariant under affine transformations of the underground data (timeseries): the visibility criteria invariant under rescaling of both horizontal and vertical axis, as well as under horizontal and vertical translations.

(iv)'Lossy': some information regarding the time series is inevitably lost in the mapping from the fact that the network structure is completely determined in the (binary) adjacency matrix. As mentioned in [3], i.e., two periodic series with the same period as $T_1 = \cdots, 3, 1, 3, 1, \cdots$ and $T_2 = \cdots, 3, 2, 3, 2, \cdots$ would have the same visibility graph, albeit being quantitatively different.

**Remark**: One straightforward question is: what does the visibility algorithm stand for? Which analysis properties exist there for a visibility graph? These questions are answered in [3]. And some extended versions of VG, such as horizontal visibility graph algorithm [3], Limited Penetrable visibility graph algorithm [17], Windowed VG [4], Dynamical VG [4], etc., are proposed during these several years, and they had been used for nonlinear timeseries anlysis and data set classification successfully, and those algorithms do work for differing one dimension signals from each others in different situations. (See [3],[4],[5], to name just a few of existed literatures).

# 3. Algorithm for Constructing Planar Visibility Graph Networks (PVGN)

In this section, a new Planar Visibility Graph Networks (PVGN) is constructed based on the classical Visibility Graph algorithm. To construct the related networks for a typical two dimension signal, i.e., a 2-D time series Ts = ([$Ts_x$,$Ts_y$]) = [(1,1),(2,4),(3,2.5),(4,3)], two main steps are contained in the new algorithm. It is listed as follows.

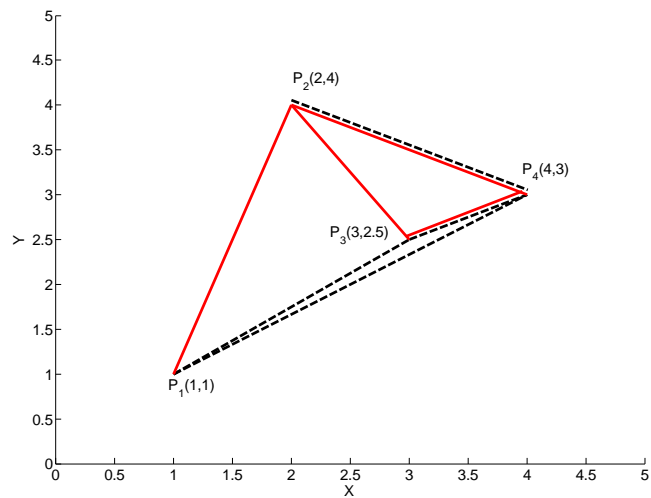

Figure 2. Sample for 2-D VG algorithm for translate 2-D timeseries into networks, where Ts=[(1,1),(2,4),(3,2.5),(4,3)], the solid line and the dash line indicates the link of x and y directions, respectively.

**Step 1: create uni-directional VG adjacent matrix through abscissa and ordinate directions;**

      I. X-direction case:

      1) Sortting $Ts_y$ according to their abscissa's order: $(x_1,x_2,x_3,x_4)$ from left to right as $Ts_y$', where the sequence $Ts_y$' is

to be applied on classical VG algorithm;

    2) Deal with standard VG processing on related $Ts_y$';

    3) Output the related X-direction VG adjacent matrix;

For example, for time series Ts=[(1,1),(2,4),(3,2.5),(4,3)] illustrated in figure 2, the resortted $Ts_y$' is [1.0, 4.0, 2.5, 3.0], so the related X-direction VG adjacent matrix is

$$Adj_x = \begin{pmatrix} 0 & 1 & 0 & 0 \\ 1 & 0 & 1 & 1 \\ 0 & 1 & 0 & 1 \\ 0 & 1 & 1 & 0 \end{pmatrix}$$

    II. Y-direction case:

    1) Sortting $Ts_x$ according to their ordinate's order: $(y_1, y_2, y_3, y_4)$ from down to up as $Ts_x$',

    2) Deal with standard VG processing on related $Ts_x$';

    3) Output the related Y-direction VG adjacent matrix;

For example, for time series Ts=[(1,1),(2,4),(3,2.5),(4,3)] illustrated in figure 2, the resortted $Ts_x$' is $Ts_x$'=[1.0, 2.5, 3.0, 4.0], so the related Y-direction VG adjacent matrix is

$$Adj_y = \begin{pmatrix} 0 & 0 & 1 & 1 \\ 0 & 0 & 0 & 1 \\ 1 & 0 & 0 & 1 \\ 1 & 1 & 1 & 0 \end{pmatrix}$$

**Step 2: composite both uni-directional VG adjacent matrixs for unweighted or weighted cases by using a choose operator;**

For unweighted case, the composited unweighted adjacent matrix M can be obtained via operations listed as follows:

$$Adj_1^* = F(M),$$

where $M = (Adj_x + Adj_y) \in R^{N \times N}$, and F( • ) is an element choose operator for creating a related Boole Matrix, which satisfies:

$$F(M(i,j)) = \begin{cases} 1, & M(i,j) = 2, \\ 0, & M(i,j)) = 0. \end{cases}$$

For example, for the case illustrated in figure 2, the related unweighted adjacent matrix is:

$$Adj_1^* = F(M) = \begin{pmatrix} 0 & 0 & 0 & 0 \\ 0 & 0 & 0 & 1 \\ 0 & 0 & 0 & 1 \\ 0 & 1 & 1 & 0 \end{pmatrix},$$

For weighted case, one can just do it as $Adj_2^* = (Adj_x + Adj_y) \in R^{N \times N}$, for the case illustrated in figure 2, the related weighted adjacent matrix is:

$$Adj_2^* = \begin{pmatrix} 0 & 1 & 1 & 1 \\ 1 & 0 & 1 & 2 \\ 1 & 1 & 0 & 2 \\ 1 & 2 & 2 & 0 \end{pmatrix}.$$

**Remark:** One can also use the Boole Matrix choose operator G( • ) satisfies:

$$G(M(i,j)) = \begin{cases} 1, & M(i,j) \geq 1, \\ 0, & M(i,j)) = 0. \end{cases}$$

to simplify the weighted adjacent matrix as a binary matrix to characterize the existence of either single linke or duoble links between two nodes:

$$Adj_2^{*\prime} = \begin{pmatrix} 0 & 1 & 1 & 1 \\ 1 & 0 & 1 & 1 \\ 1 & 1 & 0 & 1 \\ 1 & 1 & 1 & 0 \end{pmatrix}.$$

**Remark:** In fact, both the weighted matrix $Adj_2^*$ and $Adj_2^{*\prime}$ are good enough for differ perodic, random and chaotic two deminsion signal from each other according our numerical experiments. We will use the definition of $Adj_2^*$ for further discussions in the next section. In matlab, the command 'logical(A)' can fulfill the function of Boole Matrix choose operation.

## 4. Comparison constructed networks from periodic, random and chaotic timeseries.

In this section, we will transfer 2-D periodic, random and chaotic timeseries into related complex networks and compare their network characteristics from each other.

a) Periodic case.

Here, we choose the periodic trajactory of Y=[sint, cost], where t is valued in [0,10], the time step is 0.002. In figure 3, the phase portrait and the related weighted adjacent matrix are illustrated in Figure 3, respectively. The related degree distribution plot is shown in Figure 4 (with an illustration of network with 500 nodes, since too many nodes can not be expressed so clearly).

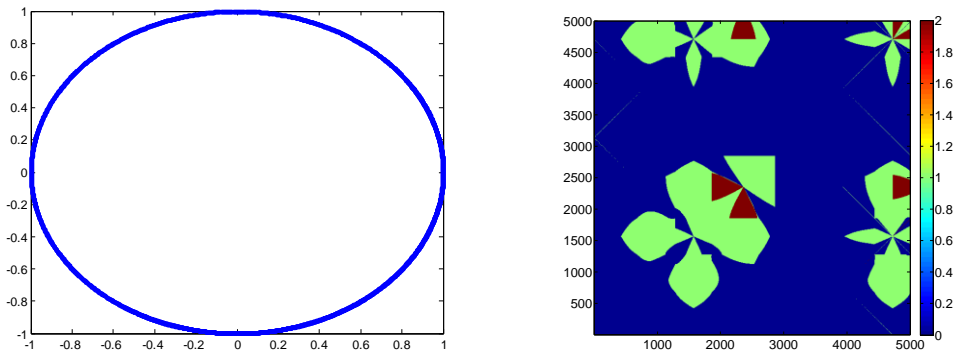

Figure 3. Left: Period trajactory of Y=[sint, cost]; Right: Related weighted adjacent matrix.

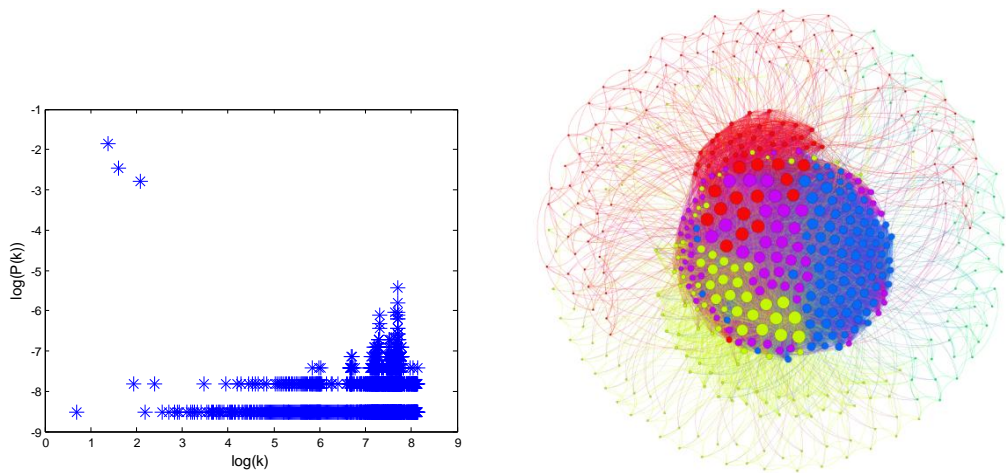

Figure 4. Left: the degree distribution of a transferred network with 5000 nodes, Right: a periodic network of Y=[sint, cost] with 500 nodes for illustration.

b) Random case.

Here, we choose the random trajactory of Y=rand(2,5000), of which each entry obeys normal distribution in [0,1]. In figure 5, the phase portrait and the related weighted adjacent matrix are illustrated in Figure 5, respectively. The related degree distribution plot is shown in Figure 6 (with an illustration of network with 500 nodes, since too many nodes cannot be expressed so clearly).

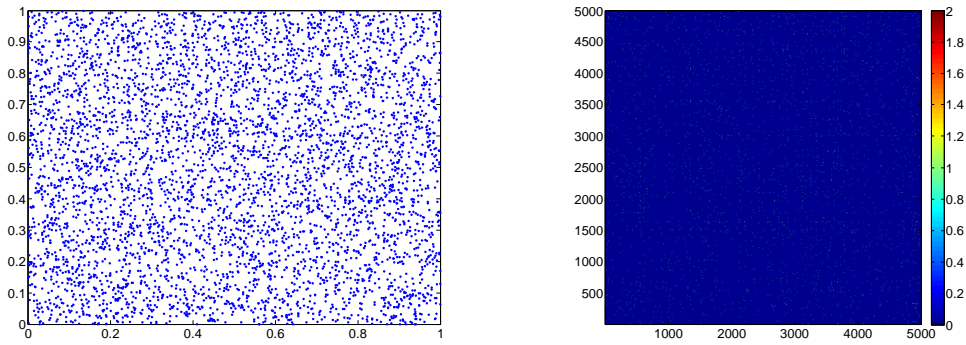

Figure 5. Left: Random trajactory of Y=random(2,5000); Right: Related weighted adjacent matrix.

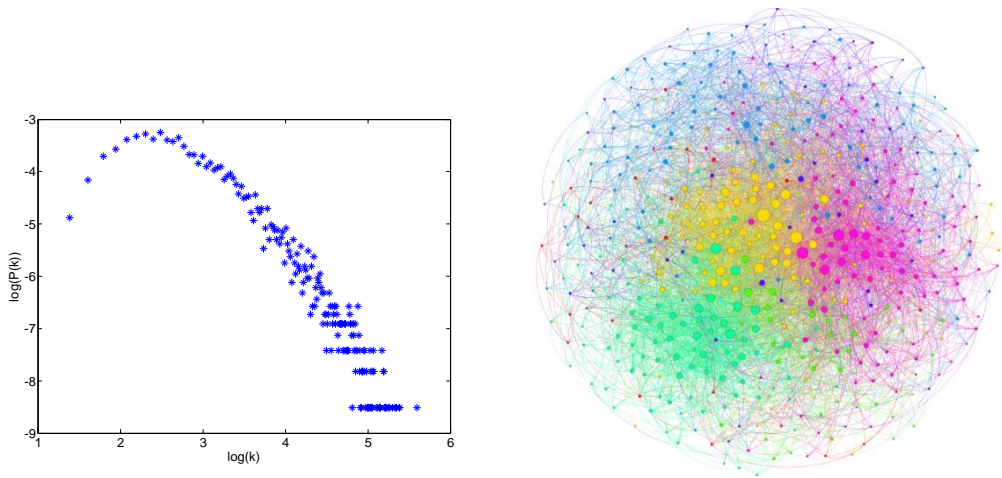

Figure 6. Left: The degree distribution P(k) of the planar visibility graph network associated with timeseries (plotted in log-log). Although the beginning of the curve approaches the result of a poisson process, the tail is clearly exponential, Right: a random network of Y=rand(2,500) with 500 nodes just for illustration.

c) Chaotic case

Here, we choose the famous chaotic trajactory of Lorenz attractor, which was firstly proposed by E. N. Lorenz in 1963. The dynamics of Lorenz attractor is described as: $\dot{x} = -c(x - y)$, $\dot{y} = ax - y - xz$, $\dot{z} = b(xy - z)$, when a=28, b=8/3, c=10, it behaves chaotic. In figure 7, the phase portrait and the related weighted adjacent matrix are illustrated, respectively. The related degree distribution plot is shown in Figure 8 (with an illustration of network with 500 nodes, since too many nodes can not be expressed so clearly).

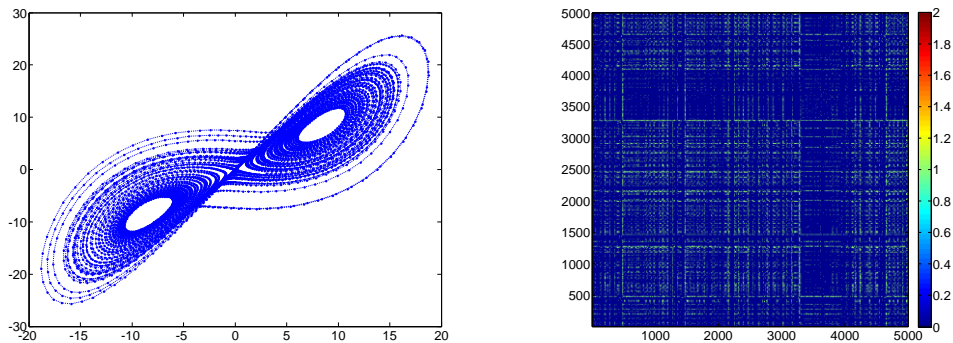

Figure 7. Left: Chaotic trajactory of Y=x(:,[2,3]); Right: Related weighted adjacent matrix.

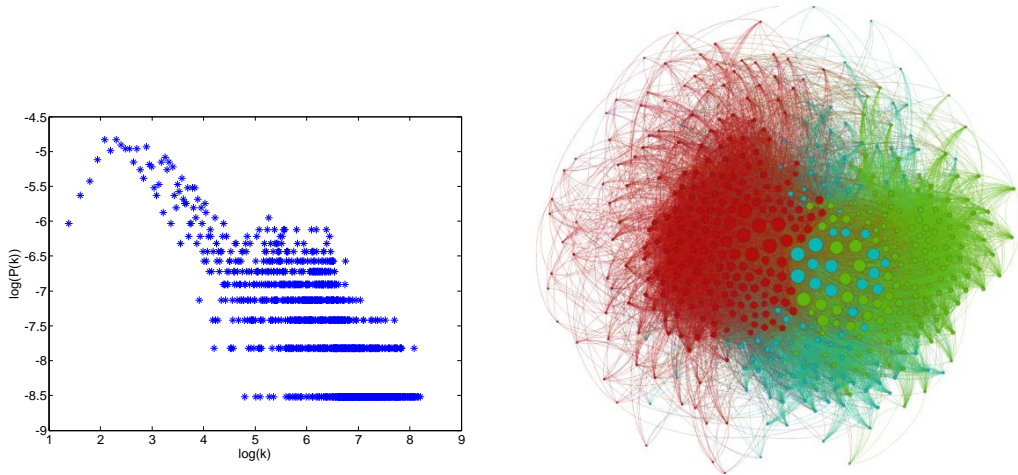

Figure 8. Left: The degree distribution P(k) of the planar visibility graph networks associated with timeseries (plotted in log-log). Although the beginning of the curve approaches the result of a poisson process, the tail is clearly piecewise exponential which is apparently different from that of periodic and random cases. Right: A chaotic network of Lorenz attractor with 500 nodes just for illustration.

d) Characteristic comparison of associated networks.

In Table 1, one can compare different networks constructed from timeseries with the same length. All characteristics are calculated through 20 runs (for chaotic series created from different initial conditions, for periodic or random series choosed from different parts of a long sequence). It can be seen that, the average degree decreases sharply from periodic, chaotic, to random. The characteristic path length(CPL) and modularity parameters both increase along with this direction, but the clustering coefficient decreases slightly at the same time. It means that one can differ three kinds of timeseries through some of network characteristics in real applications for data classfication or pattern identifications, such as ECG diagnosed from two channel ECG signals, etc.

Table 1. Differences of the related networks of aformentioned TS with length 5000. (**20 runs**)

| TS type | Average Degree | CPL | D | C | Degree Distribution exponent | Modularity |
|---|---|---|---|---|---|---|
| **Periodic** | 1179 | 2.0122 | 4 | 0.5818 | N/A | 0.143 |
| **Chaotic** | 481 | 2.0413 | 4 | 0.3689 | Piecewise Power Law exponents: Gamma = -0.000782 | 0.234 |
| **Random** | 31 | 3.1270 | 5 | 0.3342 | Power Law exponents: Gamma = -0.0265 | 0.388 |
| *Trends* | ↓ | ↑ | N/A | ↓ | N/A | ↑ |

## 5. Brief conclusions.

In summary, an algorithm that converts two dimensional timeseries into graphs is presented in this paper. The structure of the time series is conserved in the graph topology: periodic series, random series, chaotic series into different kinds of associated complex networks. With the new algorithm proposed for characterizing two dimensional timeseries, a natural bridge between complex networks theory and high dimensional timeseries analysis has now been built, which has potential usage in pattern identification and data classifications. It is a great challenge to find effective methods for characterizing the structure of time series (and the processes that generated those series) using the powerful tools of complex networks theory. Recent works have shown that the visibility graphs algorithms and its modification versions inherit several degrees of correlations from their associated series, and therefore such networks theoretical characterization seems in principle possible. However, as mentioned in [3], both the mathematical grounding of this promising theory and its applications are in its infancy, we will try to extend this new method to higher dimensional space in the future and apply it on some urgent applications.

**Acknowledgements:** This work is partly supported by NSFC under grant number No. 61203159(LJ), Key Project of Hubei Education Department under grant number No. No. D20141605(LJ), and the national inovation project for undergratuates of China under grant number No. 201210459024 (Q-Q Li).